\documentclass[preprint]{aastex}

\shorttitle{Near-Infrared monitoring of KH 15D}
\shortauthors{Kusakabe et al.}

\begin{document}

\title{NEAR-INFRARED PHOTOMETRIC MONITORING
OF A PRE-MAIN-SEQUENCE OBJECT KH 15D}

\author{Nobuhiko Kusakabe\altaffilmark{1}, Motohide Tamura\altaffilmark{1, 2}, 
Yasushi Nakajima\altaffilmark{2}}
\author{Ryo Kandori\altaffilmark{2}, Akika Ishihara\altaffilmark{2}, 
Tetsuya Nagata\altaffilmark{3}}
\author{Takahiro Nagayama\altaffilmark{3}, Shogo Nishiyama\altaffilmark{4},
Daisuke Baba\altaffilmark{4}}
\author{Shuji Sato\altaffilmark{4}, Koji Sugitani\altaffilmark{5}, 
Edwin L. Turner\altaffilmark{6}}
\author{Lyu Abe\altaffilmark{2}, Hiroshi Kimura\altaffilmark{7}, Tetsuo Yamamoto\altaffilmark{7}}

% 2005.7.25,26, version 4.5 on July 27, 4.26 on July 28 
% 5.2 on July 29 
% 6.1, 6.2, 6.3, 6.4 after Kusakabe major change in fig1, on July 31
% 6.5 some correction after circulation August 1 (Baba, Sugitani)
% 6.7 fro 6.5p after Nagata comment August 4
% rev1 after referee comment

\altaffiltext{1}{Department of Astronomical Science, Graduate University for 
Advanced Studies (Sokendai), Osawa, Mitaka, Tokyo 181-8588, Japan;
kusakabe@optik.mtk.nao.ac.jp.}
\altaffiltext{2}{National Astronomical Observatory of Japan, Osawa, Mitaka, Tokyo 
181-8588, Japan; hide@subaru.naoj.org.}
\altaffiltext{3}{Department of Astronomy, Kyoto University, Sakyo-ku, 
Kyoto 606-8502, Japan.}
\altaffiltext{4}{Department of Astrophysics, Faculty of Sciences, Nagoya University, 
Chikusa-ku, Nagoya 464-8602, Japan.}
\altaffiltext{5}{Graduate School of Natural Sciences, Nagoya City University, Mizuho, 
Nagoya 467-8501, Japan.}
\altaffiltext{6}{Princeton University Observatory, Peyton Hall, Princeton,
NJ 08544.}
\altaffiltext{7}{Institute of Low Temperature Science,
Hokkaido University, Sapporo 060-0819, Japan.}

\begin{abstract}
An extensive photometric monitoring of KH 15D, 
an enigmatic variable in the young star cluster 
NGC 2264, has been conducted. Simultaneous and accurate
near-infrared (\textit{JHKs}-bands) 
photometry is presented between 2003 December and 2005 March covering most of the 
variable phase. The infrared variability is characterized by large-amplitude and 
long-lasting eclipse, as observed at optical. 
The period of variability is 
$48.3 \pm 0.2$ days, the maximum photometric amplitude of variability is 
$\sim$4.2 mag, and
the eclipse duration is $\sim$0.5 in phase units. 
These are consistent with the most recent period, amplitude, and
duration at optical.
The blueing of the \textit{J-H} color ($\sim$0.16 mag) during the
eclipse, which has been suggested before, is unambiguously
confirmed; a similar blueing at \textit{H-Ks} is less clear but is probably
present at a similar level.
The overall shape of the \textit{JHKs} light curves 
is very similar to the optical one, including a fair time-symmetry and a less stable 
flux during the eclipse with a slight hump near the zero phase. Most of these 
variability features of KH 15D observed at near-infrared wavelengths can be explained 
with the recent model employing an eclipse by the inclined, precessing disk 
and an outer scattering region around a pre-main-sequence binary.
\end{abstract}

\keywords{circumstellar matter - stars: individual (KH 15D) - stars: 
pre-main-sequence - planetary systems: protoplanetary disks}

\section{INTRODUCTION}
KH 15D (V582 Mon; 6$^{\mathrm h}$41$^{\mathrm m}$10\fs 18, 9$^\circ$28$'$35\farcs 5, 
J2000) is a K6-7 
pre-main-sequence star (Agol et al. 2004; Hamilton et al. 2001) in NGC 2264 
($d$ = 760 pc), which 
shows a unique variability. This is the star \#15 in the D field of Kearns \& Herbst 
(1998). See Maffei, Ciprini, and Tosti (2005) for other identifications of this object. 
The optical variability is characterized by a large eclipse amplitude (maximum $\sim 
$4.0 mag in \textit{I}-band between 1999 and 2004; 
Barsunova, Grinin, \& Sergeev 2005; Johnson et al. 2005;
Hamilton et al. 2005) and 
a long eclipse duration, about half of its period (currently $\sim$24 days out of 
$\sim$48 days; Winn et al. 2004; Johnson et al. 2005). The eclipse duration has been 
increasing with time, by 1-2 days per year. 
A great deal of information about this 
system has been obtained from archival studies (Winn et al. 2003; 
Johnson \& Winn 2004; Johnson et al. 2005). 
The eclipses were formerly much shallower and the system was brighter overall. During 
the eclipse a slight blueing of the star's color indices (Herbst et al. 
2002; Hamilton et al. 2005), 
little or no change in spectral type (Hamilton et al. 2001), relatively large 
flux fluctuations (Herbst et al. 2002; Barsunova et al. 2005), and a dramatic increase 
in linear optical polarization (Agol et al. 2004)
are observed. 
The latter measurements suggest that a substantial fraction or all of the
light in eclipse is scattered light.

At near-infrared wavelengths, no color change or marginal blueing during the eclipse is 
reported (Knacke, Fajardo-Acosta, \& Tokunaga 2004).
Molecular hydrogen emission at 2.12 $\mu {\rm m}$ presumably 
associated with the mass outflow from this source has been detected (Deming et al. 
2004; Tokunaga et al. 2004), which can be circumstantial evidence for a disk
around the central star(s).
Recent spectroscopic monitoring has revealed that the system 
is in fact a single-lined spectroscopic binary with the same period of the eclipse 
(Johnson et al. 2004). These remarkable properties are now explained by the theory 
that a binary star is gradually occulted by an inclined and precessing circumbinary 
disk (Winn et al. 2004) or narrow ring (Chiang \& Murray-Clay 2004).

In spite of the recent intensive interests in this unusual object, a very limited 
amount of infrared monitoring is reported, which must be indispensable to 
understanding the occulting material, the dust extinction properties, and 
the contribution 
of dust thermal emission, if any. 
In order to better understand the enigmatic 
variability of this source, we have started a long-term monitoring of KH 15D at 
\textit{JHKs} simultaneously from 2003 December. In this $Letter$, we present the 
magnitudes and colors for the first 16 months (54 independent data points), 
which covers most of its variable phase ($\phi$) and serves as 
the most extensive
near-infrared monitoring data of KH 15D to date, 
and interpret the data based on the 
disk-eclipse model for this system.

\section{OBSERVATIONS}
We carried out imaging observations of the field
centered around KH 15D in the near-infrared 
bands \textit{J} ($\lambda_{c}$ = 1.25 $\mu {\rm m}$), \textit{H} (1.63 $\mu {\rm m}$), 
and 
\textit{Ks} (2.14 $\mu {\rm m}$) simultaneously. 
The observations were made during 2003 December and 
2005 March with the near-infrared camera SIRIUS  
on the IRSF 1.4-m telescope at 
Sutherland, South Africa. The camera is equipped with three 1024$\times$1024-pixel 
HgCdTe (HAWAII) arrays. Two dichroic mirrors enable simultaneous 
observations in the three bands. Details of the camera are described in Nagashima 
et al. (1999) and Nagayama et al. (2003). The image scale of the array is 
0 \farcs 45 pixel$^{-1}$, giving a field of view of $7\farcm 7\times 7\farcm 7$.

We measured 54 independent data points during this period. We obtained 90 dithered 
frames with typical exposure time of 10 seconds, resulting in the total integration 
time of 900 seconds for each data point. 
Typical seeing was 1\farcs 4 (FWHM), ranging from 1$''$ to 2$''$, 
in the \textit{Ks} band. 
The standard star No. 9116 in the faint infrared standard star catalog of 
Persson et al. (1998) was observed for the photometric calibrations.

We used NOAO IRAF 
software package to reduce 
the data. We applied the standard procedures of near-infrared array image reduction, 
including dark current subtraction, sky subtraction and flat fielding. 
See Nakajima et al. (2005) for the details of the SIRIUS image reductions.
Identification and photometry of point sources were performed by using the DAOFIND 
and PHOT tasks in IRAF, respectively. 
The aperture radius for the photometry was 3 pixels (1\farcs 35).

\section{RESULTS}
Fig. 1 shows a \textit{JHKs} composite color image of the observed region 
including KH 15D. 
The field also includes the famous infrared 
YSO, NGC 2264 IRS1 (Allen 1972) and the top of the Cone Nebula.
Note that KH 15D, situated at $\sim 50''$ to 
the south of IRS1, is also somewhat affected by the nebula associated with IRS1.  
In order to accurately 
measure the photometric variability of KH 15D, we employed relative photometry within 
the field. 
First we checked the variability of each source in the field compared with 
the median of all the sources among various nights. 
Then only the sources whose rms errors of magnitudes 
are less than 3$\sigma$ are selected. This process was repeated three times. Finally six 
sources in the field whose non-variability were confirmed with the above processes 
were selected (Fig. 1).

Fig. 2 shows the \textit{JHKs} light curves of KH 15D after calibrating with the 
photometric standard star (The data are in Table 1; electronic form only). The 
periodicity seen at optical wavelengths is clear even in this figure. We determined 
the periods of the \textit{JHKs} light curves to be the same, $48.3 \pm 0.2$ days. 
This is also identical to the most recent optical period.

Phased light curves using the optical period of 48.36 days 
(Herbst et al. 2002) are shown in Fig. 3. The overall 
shape of each light curve is very similar to the optical one:

(1) The light curves at \textit{JHKs} show a fairly good symmetry with time, though 
the detailed shape has some asymmetry. For example, the slopes of the first decrease 
or increase ($\phi$ = -0.25 -- -0.15 and 0.15 -- 0.25) are slightly different with each 
other. Similar changes in slopes are observed at optical (Herbst et al. 2002; Barsunova et 
al. 2004). Note that the beginning and ending of the eclipse is not as abrupt as seen 
in the 2001-2002 optical data (Herbst et al. 2002).

(2) The maximum photometric variability amplitudes are nearly 
identical at \textit{JHKs}, $\sim$4.2 mag. 
These values are consistent with the optical amplitude of 
$\sim$4.0 mag (in \textit{I}-band between 1999 and 2004, after correcting for the 
monotonous flux decrease trend; Barsunova et al. 2005; Johnson et al. 2005). 
The average amplitudes at \textit{JHKs} are $\sim$3.6 mag. 
Note that the large 
fluctuations during the eclipse make an accurate comparison of the values between 
infrared and optical difficult.

(3) The flux during the eclipse is less stable, and there is a clear flux hump near 
$\phi$ = 0. The hump at \textit{JHKs} is at a level of 0.5 mag and continues about 
$\sim$4 days. This hump is also observed at optical (Herbst et al. 2002; Barsunova 
et al. 2005). Two minima occur at $\phi$ $\sim$ $\pm$0.1; 
the minimum at $\phi$ = +0.1 appears deepest.

(4) The duration of the eclipse in 2003-2005 (0.52 in phase units at FWHM) is larger by 
30\% than the optical duration in 2001-2002 (0.4; Herbst et al. 2002),
and is rather consistent with the most recent (2003-2004) optical data (0.5; Hamilton et al. 2005). 
There is no difference of the duration phase among \textit{JHKs}.

Fig. 4 shows the phased color curves of KH 15D (\textit{J-H}, \textit{H-Ks}). Outside 
of the eclipse the \textit{J-H} and \textit{H-Ks} colors of KH 15D are constant, (0.68 
$\pm$ 0.01) and (0.28 $\pm$ 0.02) in magnitudes, respectively, which match those of 
the K7-type T Tauri stars (e.g., V410 Tau) with a slight reddening 
($A_{\sl V}$ $\sim$ 1 mag).
Knacke et al. (2004) derived the same spectral type from the average non-eclipse 
colors of \textit{J-H} = 0.67 and \textit{H-K} = 0.18. Note that the IRSF/SIRIUS color 
system is almost identical to the MKO system.
The derived spectral type is consistent with the optical one (Hamilton et al. 2001;
Agol et al. 2004).

Most striking is the clear change of infrared colors during the eclipse. As seen in 
Fig. 4, the blueing of \textit{J-H} color (${\Delta}m$ = 0.16 mag) during the transit is 
unambiguously confirmed in our observations. Such blueing has been first reported 
at optical (Herbst et al. 2002; see Hamilton et al. 2005
for the most recent optical data) 
but only marginally suggested at near-infrared by 
Knacke et al. (2004). The blueing of \textit{H-Ks} color is less clear but seems to 
be at a similar level (${\Delta}m$ $\sim$ 0.1 mag). 
The average \textit{J-H} and \textit{H-Ks} 
colors during the eclipse are (0.52 $\pm$ 0.05) and (0.18 $\pm$ 0.12)
in magnitudes, respectively.

No positional shift of KH 15D is observed between the on- and off-eclipse periods,
which suggests that a contamination by an interloper is unlikely the cause
of the blueing. In order to crosscheck this, we have considered
possible blue sources. 
Such a source must have a color of \textit{J-H} 
$<$ 0.52 (= 0.68 - 0.16) mag to explain the observed blueing. 
Since the visual extinction 
at the location of KH15D is larger than $A_{\sl V}$ = 10 mag or \textit{J-H} = 1.1 mag 
(Simon \& Dahm 2005),
the intrinsic color must be \textit{J-H}  $<$ -0.58 mag.
No foreground or background star or background galaxy 
with such a blue intrinsic color is expected 
to interlope our 1\farcs 35 aperture radius.

\section{DISCUSSION}

The most plausible model of these enigmatic variable features of KH 15D is the theory 
that a binary star in a mutual orbit with high eccentricity is gradually occulted 
by an inclined and precessing circumbinary disk (Winn et al. 2004) or narrow ring 
(Chiang \& Murray-Clay 2004). The existence of such a companion has been recently 
confirmed by radial velocity measurements (Johnson et al. 2004); the orbital 
parameters agree well with the prediction by Winn et al. (2004).
The long-term change of the variability characteristics revealed by
archival studies (Winn et al. 2003; Johnson \& Winn 2004; 
Johnson et al. 2005) is also explained with the same idea.
Today, only one component of the binary is visible (we
refer to this component as Star A), with the other component (Star
B) being entirely hidden behind the disk.
By employing this theory, we consider the near-infrared and optical 
features that are explainable as follows:

(a)	Both the large variability amplitude and the long-lasting periodic eclipse 
that are almost independent of the observed wavelengths (\textit{VRIJHKs}) are well 
explained with a gradual occultation by a knife edge screen. In this case, the screen 
is either a circumbinary disk or ring, and the disk dust size responsible for the screening 
must be much larger than the observed wavelengths ($\gg$ 2 $\mu {\rm m}$). The rough 
time-symmetry of the light curve is also explained with this theory. The detailed 
asymmetric features during the eclipse are probably due to the fine structures and 
kinematics of the disk or ring, which needs future detailed modeling.

(b)	The near-infrared color blueing as well as the optical one
indicates that the eclipse is not due to the disk dust absorption
which causes ``reddening''. 
As described below, the blueing can also be 
explained by the eclisping disk model with an outer scattering region. 
This is consistent with the increase 
of optical polarization during the eclipse which suggests that a large fraction of 
the light in eclipse is scattered light (Agol et al. 2004). 
% the abobe two have been changed, as suggested by the referee comment No11.  

The polarization data 
suggest that the scattering region is not completely obscured by the occulting 
material. The dust in this scattering region, which is distinct
from the screening dust mentioned above, must be responsible for
both the color blueing and the polarization.
%Since the blueing is observed not only at \textit{JHKs} but also at 
%\textit{VRI} (Herbst et al. 2002), the dust size responsible for the scattering, which 
%is distinct from the screening dust mentioned above, must be small ($a<\lambda /2\pi 
%\sim$ 0.1 $\mu {\rm m}$), comparable to the interstellar dust size. This is in contrast to 
%the conclusion by Agol et al. (2004) who considered that the eclipse scattering is 
%neutral and the scattering dust size is relatively large (5-10 $\mu {\rm m}$). 
In order to explain these, we have calculated light scattering by dust grains
in the non-occulted, scattering region. We have assumed the scattering region to be 
a semi-sphere over the eclipsing disk, the dust spatial distribution to follow
r$^{-1.5}$ (corresponding to free-falling dust), and the dust material to be
silicate (Laor \& Draine 1993), and the dust size distribution to follow
a$^{-3.5}$ with a$_{\rm min}$ = 0.5 nm and a$_{\rm max}$ = 5 $\mu {\rm m}$.
The radius of the semi-sphere is 2.6 AU with its equatorial plane
inclined by 20$^\circ$ to the line-of-sight (Winn et al. 2004).
The resultant color changes and optical polarization of the integrated
scattered light are
${\Delta}m$ = 0.16 mag for \textit{J-H} and 0.18 mag for \textit{H-Ks}, and
$p$(optical)=2\%, respectively. Although detailed modeling of the
geometry and dust grains is beyond the scope of this paper,
this simple model quantitatively well reproduces both 
our observed color changes and the polarizations by Agol et al. (2004).

Since there is no sign of \textit{H-Ks} color excesses,
%\textit{H-Ks} color also shows a (marginal) blueing, 
the possible thermal emission 
from the disk suggested from the model or the H$_2$ outflows appears insignificant 
at near-infrared.

%(c)	Weak slopes at the start and end of the eclipse are seen 
%in our infrared 2003-2005 
%data, though the number of the non-eclipse infrared data is limited. In contrast, 
%the optical 2001-2002 data has much sharper slopes. If confirmed, such ``softening'' 
%of the knife edge is either due to a time change or a less extinction, at infrared 
%than at optical, in the outer parts of the disk.

(c)	The hump near $\phi$ = 0 is probably due to some flux contribution of the unseen 
star (Star B) of the binary in the Winn's model because the Star B is nearest to the 
occulting edge at the middle of the eclipse. The amplitude is $\sim$0.5 mag at 
\textit{JHKs}, almost identical to the amplitude at \textit{I} (in 2002-2003, Johnson 
et al. 2005). 
The optical color tends to be slightly bluer near this hump 
compared with other eclipsing color (Hamilton et al. 2005), 
which might support the interpretation 
that some additional light (from Star B) increases at this time. 
Since such a change of our near-infrared color near this hump is not
conclusive, a more accurate and intensive monitoring during this phase is necessary.
% the abobe para has been changed, as suggested by the referee science comment No3.  

\section{CONCLUSIONS}
We have conducted \textit{JHKs} monitoring of KH 15D in NGC 2264 between 2003 December 
and 2005 March. The main conclusions of this paper are as follows:

(1)	The \textit{JHKs} light curves are very similar to the optical light curve, 
which show a fairly good time-symmetry and a slight flux hump near the zero phase.

(2)	The \textit{JHKs} period is the same as the optical period, 48.3 days.

(3)	The maximum \textit{JHKs} variability amplitude at near-infrared is 
$\sim$4.2 mag.
%, while their average value is $\sim$3.6 mag. 
%These are also consistent 
%with the optical values.

(4)	The eclipse durations  are identical among \textit{JHKs} (0.52 in phase units
from 2003 to 2005).

(5)	The blueing of \textit{J-H} color (${\Delta}m$ = 0.16 mag) during the transit is 
unambiguously confirmed, while the blueing of \textit{H-Ks} color appears at a 
similar level.

(6)	These near-infrared variability features can be explained with the model 
employing an eclipse by the inclined and precessing disk or ring around a 
pre-main-sequence binary. 
The dust in the disk causing the eclipse is very large in size (a $\gg$ 2 $\mu {\rm m}$), 
therefore the
eclipse is independent of wavelengths, 
while the dust responsible for the scattered flux during the eclipse is 
outside of the disk and smaller in size (a$_{\rm max}$ $\sim$ 5 $\mu {\rm m}$),
causing the color-blueing and the polarization only during this phase.

\acknowledgments

We are grateful to Josh Winn and Chie Nagashima for helpful comments.
We thank T. Tanabe, R. Kadowaki, and Y. Haba for their help in observations.
We acknowledge M. Fukagawa, T. Naoi, and N. Kaneyasu for 
helpful discussions and assistance. 
We also thank the referee John Stauffer for his useful suggestions.
MT, TN, HK, TY and SS are supported by Grants-in-Aid from the Ministry
of Education, Culture, Sports, Science, and Technology.

%Table 1

\begin{deluxetable}{ccccccc}[!htb]
\tablecolumns{7}
\tablewidth{0pc}
\tablecaption{Near infrared photometry of KH 15D.}
\tablehead{
\colhead{JD(day)} & \colhead{\textit{J}} & \colhead{\textit{J} err} & \colhead{\textit{J-H}} & \colhead{\textit{J-H} err} & \colhead{\textit{H-Ks}} & \colhead{\textit{H-Ks} err}}
\startdata
2452989.42 & 16.88 & 0.03 & 0.51 & 0.04 & 0.25 & 0.02 \\
2453002.38 & 13.42 & 0.02 & 0.65 & 0.03 & 0.27 & 0.01 \\
2453066.41 & 16.51 & 0.03 & 0.54 & 0.04 & 0.16 & 0.04 \\
2453075.28 & 17.16 & 0.02 & 0.47 & 0.03 & 0.13 & 0.02 \\
2453078.30 & 17.07 & 0.02 & 0.52 & 0.03 & 0.13 & 0.03 \\
2453291.59 & 13.42 & 0.02 & 0.68 & 0.02 & 0.27 & 0.02 \\
2453292.59 & 13.41 & 0.01 & 0.67 & 0.01 & 0.27 & 0.02 \\
2453293.59 & 13.41 & 0.01 & 0.67 & 0.01 & 0.30 & 0.01 \\
2453302.61 & 13.45 & 0.00 & 0.69 & 0.01 & 0.28 & 0.02 \\
2453306.61 & 14.53 & 0.00 & 0.67 & 0.01 & 0.34 & 0.02 \\
2453307.67 & 16.31 & 0.02 & 0.56 & 0.03 & 0.26 & 0.04 \\
2453309.55 & 16.73 & 0.01 & 0.55 & 0.02 & 0.28 & 0.03 \\
2453312.55 & 17.19 & 0.01 & 0.54 & 0.01 & 0.04 & 0.03 \\
2453313.59 & 17.35 & 0.01 & 0.58 & 0.02 & 0.16 & 0.04 \\
2453315.57 & 17.34 & 0.01 & 0.49 & 0.02 & 0.13 & 0.03 \\
2453317.60 & 17.29 & 0.01 & 0.54 & 0.02 & 0.29 & 0.02 \\
2453318.46 & 17.16 & 0.01 & 0.53 & 0.02 & 0.17 & 0.03 \\
2453318.53 & 17.20 & 0.01 & 0.61 & 0.02 & 0.25 & 0.02 \\
2453318.60 & 17.20 & 0.01 & 0.60 & 0.01 & 0.24 & 0.01 \\
2453322.51 & 17.26 & 0.01 & 0.40 & 0.01 & -0.01 & 0.03 \\
2453322.54 & 17.33 & 0.01 & 0.54 & 0.02 & 0.07 & 0.03 \\
2453384.42 & 13.56 & 0.02 & 0.67 & 0.02 & 0.28 & 0.01 \\
2453386.36 & 13.49 & 0.02 & 0.68 & 0.02 & 0.25 & 0.03 \\
2453388.41 & 13.45 & 0.01 & 0.67 & 0.02 & 0.28 & 0.02 \\
2453398.21 & 13.46 & 0.04 & 0.67 & 0.04 & 0.26 & 0.02 \\
2453400.22 & 13.53 & 0.01 & 0.67 & 0.02 & 0.27 & 0.02 \\
2453401.22 & 13.61 & 0.01 & 0.69 & 0.02 & 0.28 & 0.02 \\
2453405.25 & 16.63 & 0.02 & 0.61 & 0.03 & 0.06 & 0.04 \\
2453405.28 & 16.64 & 0.02 & 0.54 & 0.03 & 0.25 & 0.03 \\
2453405.35 & 16.63 & 0.02 & 0.56 & 0.03 & 0.18 & 0.03 \\
2453406.28 & 16.81 & 0.01 & 0.54 & 0.02 & 0.28 & 0.03 \\
2453406.36 & 16.83 & 0.01 & 0.54 & 0.02 & 0.16 & 0.02 \\
2453407.21 & 16.95 & 0.01 & 0.55 & 0.02 & 0.22 & 0.02 \\
2453409.37 & 17.22 & 0.02 & 0.45 & 0.03 & 0.24 & 0.02 \\
2453411.22 & 17.38 & 0.02 & 0.49 & 0.02 & -0.06 & 0.02 \\
2453415.26 & 17.25 & 0.02 & 0.54 & 0.03 & 0.31 & 0.03 \\
2453417.24 & 17.25 & 0.01 & 0.53 & 0.02 & 0.35 & 0.02 \\
2453421.26 & 17.48 & 0.02 & 0.55 & 0.04 & 0.05 & 0.03 \\
2453424.22 & 17.06 & 0.02 & 0.52 & 0.02 & 0.42 & 0.03 \\
2453426.27 & 16.70 & 0.01 & 0.55 & 0.01 & 0.19 & 0.03 \\
2453427.19 & 16.54 & 0.02 & 0.60 & 0.03 & 0.27 & 0.04 \\
2453427.32 & 16.52 & 0.02 & 0.62 & 0.03 & 0.30 & 0.04 \\
2453428.15 & 16.36 & 0.02 & 0.62 & 0.04 & 0.33 & 0.04 \\
2453428.31 & 16.33 & 0.02 & 0.59 & 0.03 & 0.32 & 0.04 \\
2453429.26 & 14.91 & 0.01 & 0.68 & 0.03 & 0.37 & 0.03 \\
2453429.34 & 14.79 & 0.01 & 0.68 & 0.02 & 0.35 & 0.02 \\
2453430.23 & 13.97 & 0.01 & 0.67 & 0.02 & 0.30 & 0.02 \\
2453430.30 & 13.93 & 0.01 & 0.69 & 0.03 & 0.30 & 0.03 \\
2453431.21 & 13.62 & 0.02 & 0.69 & 0.03 & 0.30 & 0.03 \\
2453431.30 & 13.60 & 0.02 & 0.68 & 0.03 & 0.31 & 0.03 \\
2453432.22 & 13.54 & 0.01 & 0.69 & 0.03 & 0.30 & 0.03 \\
2453432.30 & 13.54 & 0.01 & 0.69 & 0.02 & 0.30 & 0.02 \\
2453433.23 & 13.53 & 0.01 & 0.69 & 0.03 & 0.31 & 0.03 \\
2453433.31 & 13.53 & 0.01 & 0.70 & 0.02 & 0.29 & 0.03 \\
\enddata
\end{deluxetable}

%
%fig 1
%
\begin{figure}[h]
\plotone{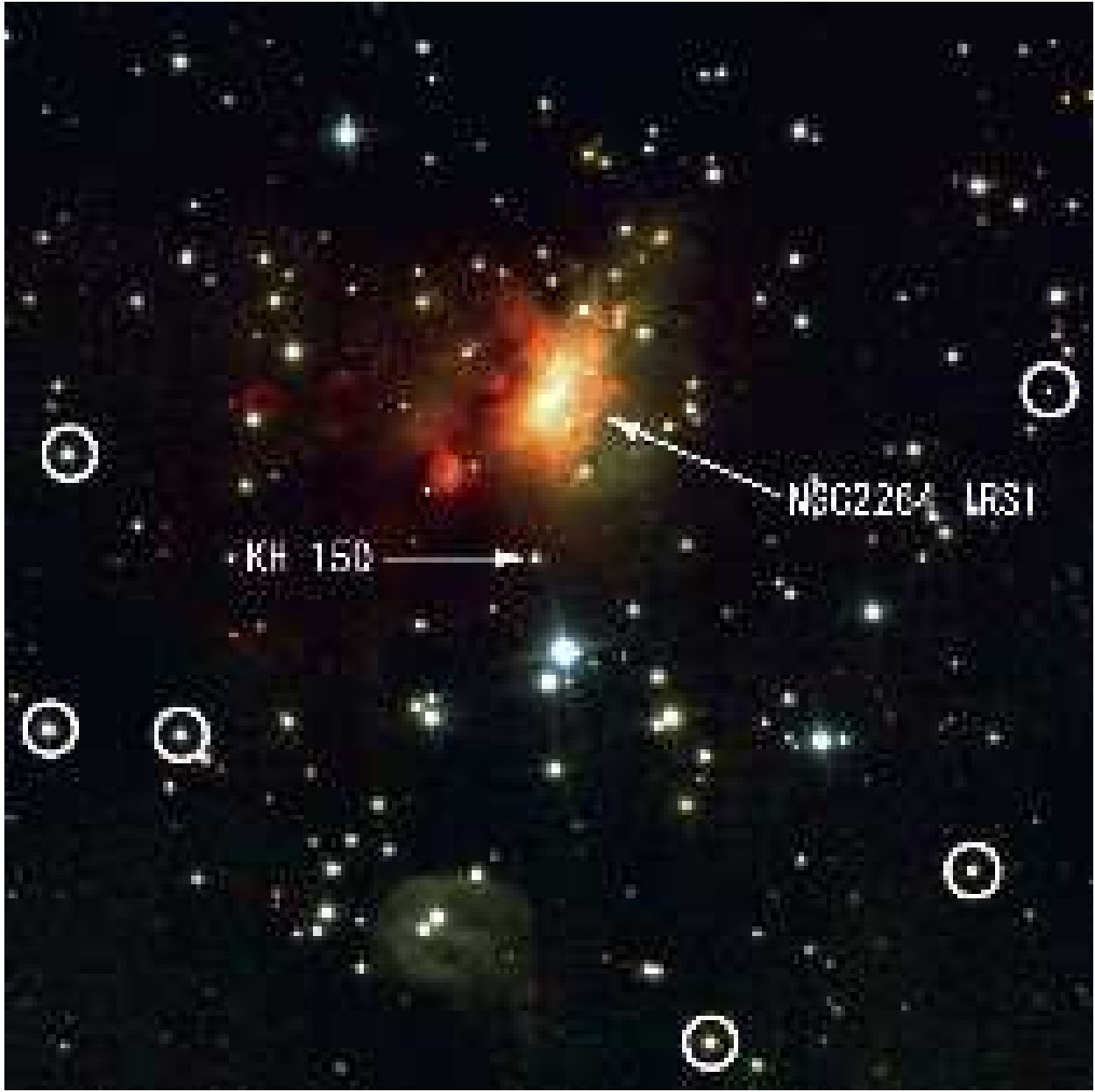}
\caption{\textit{JHKs} 
composite color image of the KH 15D region including NGC 2264 IRS1
and the top of the Cone Nebula (seen as a faint nebula to the south).
The field of view is $7'\times 7'$. North is 
top and east is to the left. 
The six stars in circles are the reference stars used for the relative 
photometry.
The image was obtained on 2004 October 23 when the system is
out of eclipse.
{\bf Note that the image quality is significantly degraded
in this preprint because of the limited file size.}
\label{fig1}}
\end{figure}
%Fig. 2 Fig. 3
\begin{figure}[h]
\plotone{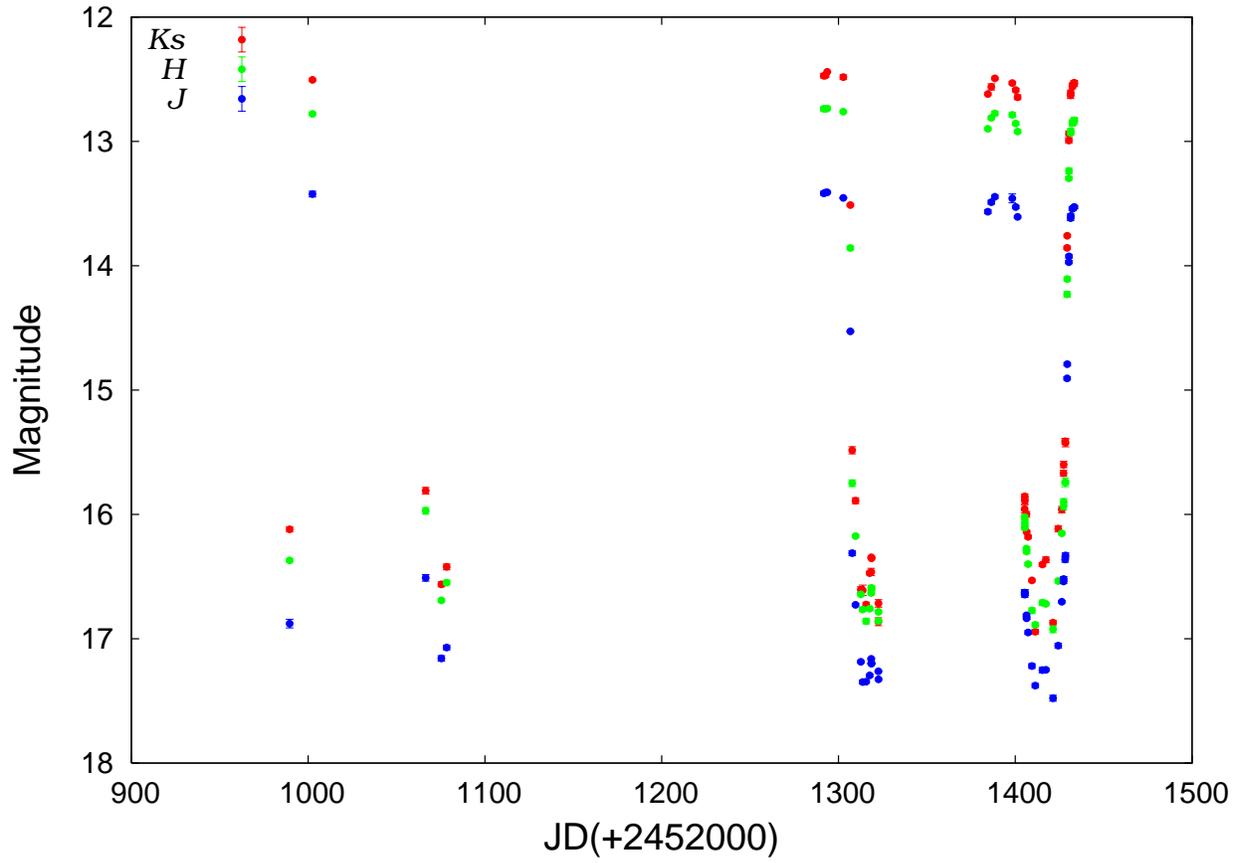}
\caption{Raw (non-phased) light curves of KH 15D (\textit{JHKs})
from 2003 December to 2005 March. \label{fig2}}
\end{figure}
\begin{figure}[h]
\plotone{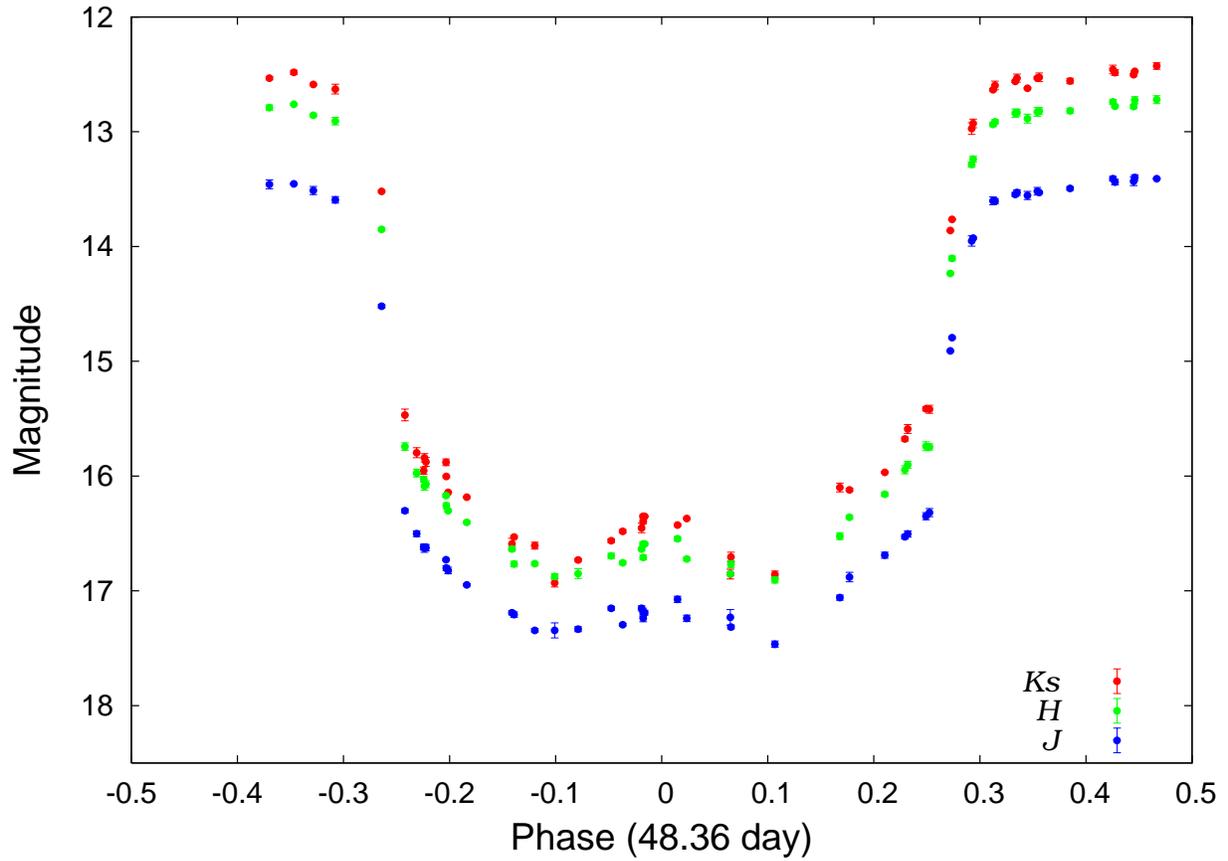}
\caption{Phased light curves of KH 15D (\textit{JHKs})
from 2003 December to 2005 March.
The period of 48.36 days
is used for phasing. \label{fig3}}
\end{figure}
%Fig. 4
\begin{figure}[h]
\plotone{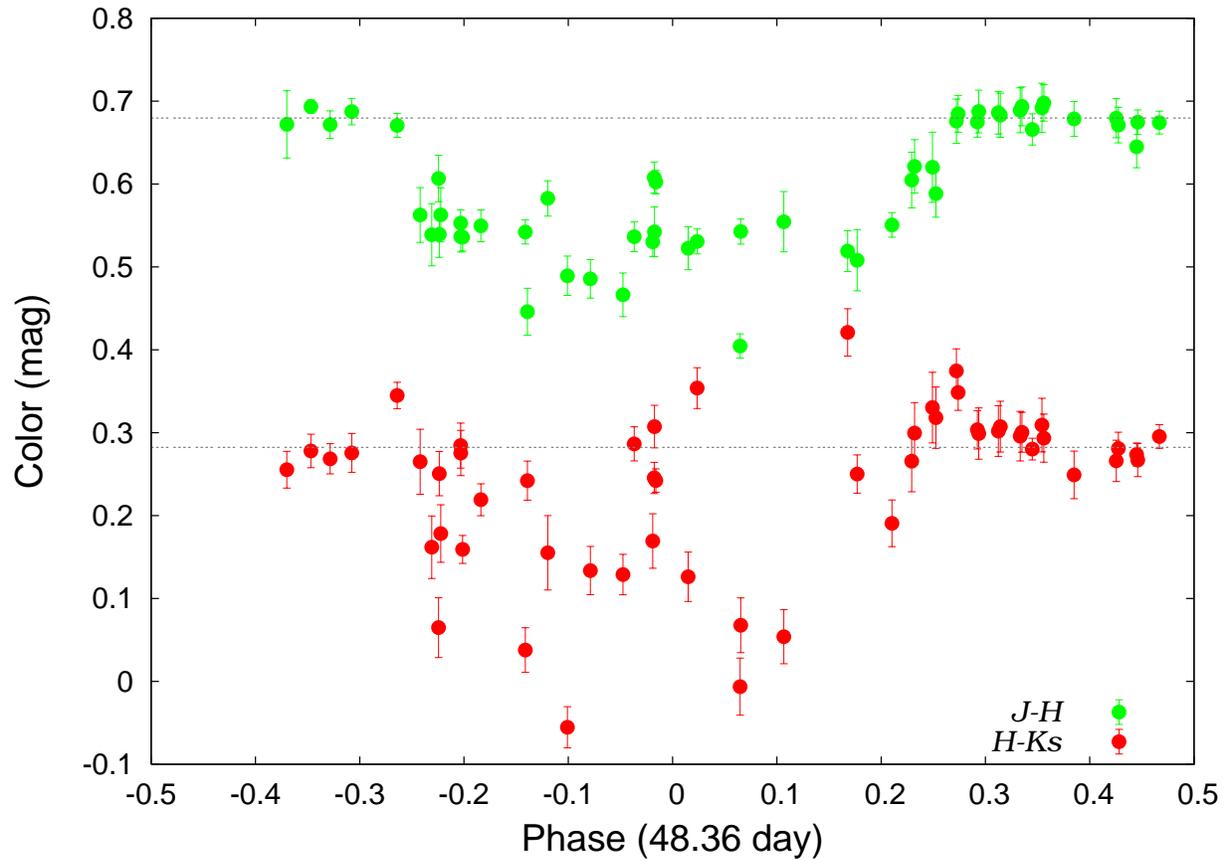}
\caption{Phased color curves of KH 15D (\textit{J-H}, \textit{H-Ks})
from 2003 December to 2005 March. The period of 48.36 days
is used for phasing. 
The dotted lines show the average colors outside of
the eclipse. \label{fig4}}
\end{figure}

\end{document}